\documentclass[prd,twocolumn]{revtex4-2}
\pdfoutput=0
\usepackage{graphicx}
\usepackage{amsmath,amssymb,amsthm,amsbsy}
\usepackage{latexsym}
\usepackage{amsfonts}
\usepackage{amssymb}

\begin{document}

\title{Extracting Dynamical Degrees of Freedom From the Quasi-Local Energy Term in the Gravitational Action}
\author{Bjoern S. Schmekel}
\affiliation{Department of Physics, College of Studies for Foreign Diploma Recipients at the University of Hamburg, 20355 Hamburg, Germany}
\email{bss28@cornell.edu}

\begin{abstract}
It is shown that under proper conditions in an appropriate coordinate system with a suitable time slicing the Hamiltonian and the Einstein-Hilbert action 
including all necessary boundary terms can be written on shell in terms of the Brown-York quasi-local energy in the absence of matter. 
If matter is present the non-vanishing bulk term only consists of stress-energy.  It is argued that the dynamical content of general relativity is stored in the quasi-local energy term. 
The results underscore the interpretation of the Brown-York quasi-local energy as the field energy of the gravitational field plus stress-energy.  
As an application we derive uncertainty relations of the time-energy kind which may be useful in the understanding of gravity induced quantum state reduction
and the more conventional kind for conjugate variables. 
The latter is computed for a modified Vaidya metric which may be used in the investigation of black hole radiance. The boundary terms expressed as quasi-local energy
cancel second derivatives in the action leaving only a square of a first derivative term in the chosen gauge which is desirable for a quantization of the action.
\end{abstract}

\maketitle

\section{Introduction}
Quantizing gravity is one of the outstanding challenges in theoretical physics. Traditional perturbative approaches lead to non-renormalizable theories
which need an infinite number of counter-terms to cancel the emerging divergences. In contrast the divergences arising in quantum field theory can be avoided
by integrating the Lagrangian density only up to a certain cut-off which can be absorbed by introducing the renormalized mass and charge. This, however, 
does not work in quantum gravity, and to make matters worse due to the equivalence principle inertial mass and gravitational mass, which is acting as the charge
of gravity, are identical.

On a more conceptual and less technical level the problems can be traced back to the metric tensor which acts both as the dynamical field of gravity and the
background spacetime. The latter also allows for an infinite amount of diffeomorphisms which corresponds to the ordinary gauge freedom in quantum field
theory. While this is a known problem which can be dealt with and which is also present in quantum field theory the role of time differs fundamentally in
the two theories. Whereas time is just a parameter in field theory labeling a sequence of events in general relativity there is no real distinction between
time and space. The usual quantization prescriptions therefore seem to rely at least on a 3+1 split of spacetime which requires a rather arbitrary choice
of the direction of time. 

While it is not the purpose of the present work to give a full working theory of quantum gravity it is attempted to rewrite the Hamiltonian and the classical action principle in a form
which mitigates some of the problems mentioned above hiding the dynamics of the theory in the boundary term as much as possible. The aim is to write all terms contributing
to the Hamiltonian or the action in terms of either stress-energy or quasi-local energy which allows for a simple interpretation of the resulting expressions. Uncertainty relations are
deduced from them. 

The paper is organized as follows. In section II the action principle generating the Einstein field equations is reviewed. The boundary term which is added
to the Einstein-Hilbert action \cite{PhysRevLett.28.1082} is of utmost importance. It gives rise to a natural definition of energy in general relativity known as the Brown-York quasi-local 
energy ("QLE") \cite{Brown:1992br} and breaks the non-distinguishability of space and time.
In fact, it is shown that in timeslicings satisfying one of several possible conditions the boundary term which is being re-expressed in terms of the QLE is the only contribution
to the action on shell apart from a contribution from stress-energy.
In section III and IV we express the Hamiltonian in geodesic gauge recovering the interpretation of the Brown-York quasi-local energy as gravitational field energy and derive an uncertainty relation for energy and time. In section V and VI we rewrite the action setting a constraint on $K_{\mu \nu} K^{\mu \nu} - (K)^2$ and compute the former explicitly for a modified Vaidya metric deriving an uncertainty relation for a field component. Alternative constraints are explored in section VII.  We conclude our results in section VIII.

\section{Action Principle and York Boundary Term}
Consider the Einstein-Hilbert action together with the York boundary term \cite{PhysRevLett.28.1082} integrated over the boundary $\partial M$ of the spacetime $M$ and a reference term $S_0$ where the integration over
$\partial M$ has been split into an integral over the time evolution $^3 B$ of a spacelike two-boundary $B$ with unit vector $n^{\mu}$ and an integration over the timeslices $\Sigma$ at
$t=t_i$ and $t=t_f$, respectively. Including the boundary term is necessary in order to derive the equations of motion from the action for finite $M$ canceling second order derivative terms in the action
\begin{eqnarray} \nonumber
S & = & \frac{1}{2 \kappa} \int_M d^4 x \sqrt{-g} \mathcal{R} + \frac{1}{\kappa} \int_{t_i}^{t_f} d^3 x \sqrt{h} K \\ 
& - & \frac{1}{\kappa} \int_{^3 B} d^3 x \sqrt{-\gamma} \Theta - S_0[\gamma_{ij}] + S_m
\label{action}
\end{eqnarray}
A summary of the used notation can be found in table \ref{notation}. 
Using the relation $\Theta = k - n_{\mu} a^{\mu}$ and
\begin{eqnarray}
\mathcal{R}  = R +  K_{\mu \nu} K^{\mu \nu} - (K)^2  - 2 \nabla_{\mu} \left ( K u^{\mu} + a^{\mu} \right )
\label{RKKK}
\end{eqnarray} we recast the action as \cite{Brown:1992br}
\begin{eqnarray}
\nonumber
S = \frac{1}{2 \kappa} \int_M d^4 x N \sqrt{h} \left [ R +  K_{\mu \nu} K^{\mu \nu} - (K)^2 \right ] \\
- \frac{1}{\kappa} \int_{^3 B} d^3 x N \sqrt{\sigma} k_1 - S_0[\gamma_{\mu \nu}] + S_m
\label{action2}
\end{eqnarray}
where $a^{\beta} \equiv u^{\rho} \nabla_{\rho} u^{\beta}$ and the index or superscript "1" denotes unreferenced quantities and the label "0" reference terms, respectively. The surface gravity is $\kappa = 8 \pi$ and $\sqrt{-g} = N \sqrt{h}$ with lapse $N$. 
We identify the last integral as a one-dimensional integral of the QLE. A derivation of these and other useful relations exploiting the Gauss-Codazzi equations can be found in the appendix of  \cite{Brown:1992br}. 

Questions whether or not there is a unique reference term $S_0$ shifting the zero-point of the QLE
are irrelevant at the classical level since any choice of $S_0$ which only depends on the induced metric $\gamma_{\mu \nu}$ of the boundary $B$ leads to the same equations of
motion. However, once the action is to be quantized, e.g. by means of a Feynman path integral prescription, this picture may change. 

We recast the bulk term exploiting relation eqn. \ref{RKKK} one more time.  Also, note that
\begin{eqnarray}
u^{\mu} u^{\nu} \mathcal{R}_{\mu \nu} = (K)^2 - K_{\mu \nu} K^{\mu \nu} + \nabla_{\mu} \left ( K u^{\mu} + a^{\mu} \right )
\end{eqnarray}
Combing these equations would give us the first of the initial value constraints \cite{Wald:1984}
\begin{eqnarray}
R -  K_{\mu \nu} K^{\mu \nu} + (K)^2  = 
2 u^{\mu} u^{\nu} G_{\mu \nu} 
\label{IVC1}
\end{eqnarray}
\begin{eqnarray}
D_b K^b_a - D_a K^b_b =  u^c h^b_a G_{bc}
\label{IVC2}
\end{eqnarray}
and they allow us to write eqn. \ref{RKKK} alternatively as
\begin{eqnarray}
\nonumber
R +  K_{\mu \nu} K^{\mu \nu} - (K)^2  = 
\\ \nonumber
\mathcal{R} + 2 u^{\mu} u^{\nu} \mathcal{R}_{\mu \nu}  + 2 K_{\mu \nu} K^{\mu \nu} - 2 (K)^2 =
\\
2  u^{\mu} u^{\nu} G_{\mu \nu}  + 2 K_{\mu \nu} K^{\mu \nu} - 2 (K)^2
\label{RKKK2}
\end{eqnarray}
using the Einstein tensor  $G_{\mu \nu} = \mathcal{R}_{\mu \nu} - \frac{1}{2} \mathcal{R} g_{\mu \nu} $.

\section{Geodesic Gauge}
From \cite{Brown:1992br} we observe that
\begin{eqnarray}
\nonumber
N \sqrt{h} \left [ K_{\mu \nu} K^{\mu \nu} - (K)^2 \right ] =\\
2 \kappa \left [ P^{ij} \dot h_{ij} - 2 P^{ij} D_i \beta_j - 2 \kappa N G_{ijkl} P^{ij} P^{kl} \right ]
\label{relcanonicalform}
\end{eqnarray}
with the canonical momentum $P^{ij}$ 
\begin{eqnarray}
P^{ij} = \frac{\delta S_{\rm cl}}{\delta h_{ij}} = ( 2 \kappa )^{-1} \sqrt{h} \left ( K h^{ij} - K^{ij} \right )
\end{eqnarray}
and inverse superspace metric
\begin{eqnarray}
G_{ijkl} = \frac{1}{2 \sqrt{h}} \left ( h_{ik} h_{jl} + h_{il} h_{jk} - h_{ij} h_{kl} \right )
\end{eqnarray}
Rewriting eqn. \ref{relcanonicalform} using the gravitational contribution to the Hamiltonian and the momentum constraint
\begin{eqnarray}
\mathcal{H} & = & 2 \kappa G_{ijkl} P^{ij} P^{kl} - (2 \kappa)^{-1} \sqrt{h} R \\
\mathcal{H}_i & = & - 2 D_j P_i^j
\end{eqnarray}
yields
\begin{eqnarray}
\nonumber
N \sqrt{h} \left [ R+ K_{\mu \nu} K^{\mu \nu} - (K)^2 \right ] = \\
2 \kappa \left [ P^{ij} \dot h_{ij} - 2 P^{ij} D_i \beta_j - N \mathcal{H} \right ]
\end{eqnarray}
The Hamiltonian $H$ follows from the action in canonical form
\begin{eqnarray}
\nonumber
S= \int_M d^4 x \left [ P^{ij} \dot h_{ij} - N \mathcal{H} - \beta^i \mathcal{H}_i \right ] \\
- \int_{^3 B} d^3 x \sqrt{\sigma} \left [ N \epsilon - \beta^i j_i \right ]
\end{eqnarray}
Assuming a generic form of $S_0$ the Hamiltonian which in classical mechanics is considered the total energy of a system can be written as \cite{Brown:1992br}
\begin{eqnarray}
\nonumber
H = \int_\Sigma d^3 x \left ( N \mathcal{H} + \beta^i \mathcal{H}_i \right ) + \int_B d^2 x \sqrt{\sigma} \left ( N \epsilon - \beta^i j_i \right ) \\
\end{eqnarray}
where $\epsilon=\left . \left ( k / \kappa \right ) \right |_0^{\rm cl}$ is the quasi-local energy surface density and $j_a= \left . -2  ( \sigma_{ak} n_l P^{kl} / \sqrt{h} ) \right |_0^{\rm cl}$ the quasi-local momentum surface density 
with ${\rm cl}$ indicating evaluation at a classical solution fo the Einstein field equations giving rise to the interpretation of the QLE as the value of the Hamiltonian generating unit proper-time translations on $^3 B$ orthogonal to $B$. 

Alternatively, we can write the Hamiltonian as
\begin{widetext}
\begin{eqnarray}
H = \int_\Sigma d^3 x \left [ \underbrace{- \frac{N \sqrt{h}}{2 \kappa} \overbrace{  \left ( R+ K_{\mu \nu} K^{\mu \nu} - (K)^2  \right ) }^{2  u^{\mu} u^{\nu} G_{\mu \nu}  + 2 K_{\mu \nu} K^{\mu \nu} - 2 (K)^2} + P^{ij} \dot h_{ij} - 2 P^{ij} D_i \beta_{j} }_{N \mathcal{H}} + \beta^i \mathcal{H}_i \right ] +
\int_B d^2 x \sqrt{\sigma} \left ( N \epsilon - \beta^i j_i \right )
\end{eqnarray}
\end{widetext}
We impose the geodesic gauge with $N=1$ and $\beta^{i}=0$. In the geodesic gauge the extrinsic curvature takes on the simple form $K_{ij} = - \frac{1}{2} \dot h_{ij}$. Together with eqn. \ref{RKKK2} we obtain
\begin{eqnarray}
H = - \frac{1}{\kappa} \int_\Sigma d^3 x \sqrt{h} u^{\mu} u^{\nu} G_{\mu \nu} + \frac{1}{\kappa} \int_B d^2 x \sqrt{\sigma} k
\end{eqnarray}
Restricting to classical solutions of the Einstein field equations we can replace the Einstein tensor by the stress-energy tensor
\begin{eqnarray}
H = - \int_\Sigma d^3 x \sqrt{h} u^{\mu} u^{\nu} T_{\mu \nu} + \int_B d^2 x \sqrt{\sigma} u_i u_j \tau^{i j}
\label{InterpretationQLE}
\end{eqnarray}
This result serves to illustrate that the quasi-local energy contains both the energy due to stress-energy and gravitational binding energy whereas the Hamiltonian as the generator of unit time translation contains only the latter. 

\section{Time-Energy Uncertainty Relation}
As an application for the Hamiltonian in geodesic gauge we derive a time-energy uncertainty relation. Since time is not an observable quantity in quantum mechanics with no operator for time existing time-energy uncertainty relations have always been somewhat controversial. The problem is directly related to the problem of time in quantum mechanics where the role of time is ambiguous and different from the purpose of position. Nevertheless it is possible to derive several meaningful time-energy uncertainty relations but the meaning of time and energy is not unambiguous and needs to be clarified carefully. For an overview of the subject cf. \cite{Busch2008,1996AmJPh..64.1451H,Mandelstam1991} and \cite{Brunetti:2002cy,Brunetti:2001mw} for examples of such time-energy uncertainty relations.
Contrast this with time in general relativity where time is part of a four-dimensional spacetime with no real distinction between time and position it is obvious that a unification of general relativity and quantum mechanics is challenging. While this is true for general relativity written in terms of the Einstein field equations this aspect is somewhat different when expressing general relativity in terms of an action principle. Even though the action in eqn. \ref{action} is absolutely covariant time enters the action as a unit vector needed to describe the boundary $\partial M$ of the necessary boundary term. While a problem in classical general relativity can be solved using the equations of motion it is almost certain that a quantum mechanical treatment will make use of an action principle or a Hamiltonian, so time would become a special quantity once the theory is quantized automatically.
For a derivation we make use of the Ehrenfest theorem
\begin{eqnarray}
\frac{d}{dt} \left \langle A \right \rangle = \frac{i}{\hbar} \left \langle \left [ H,A \right ] \right \rangle
\end{eqnarray}
Since there is no time operator we use any other operator $A$ different from the Hamiltonian which allows us to use the corresponding uncertainty relation $\Delta H \Delta A \ge \hbar / 2$. Associating the uncertainty $\Delta A= \Delta t  \frac{d}{dt} \left \langle A \right \rangle$ with the time scale $\Delta t$ gives
\begin{eqnarray}
\Delta E \Delta t \ge \frac{\hbar}{2}
\end{eqnarray}
where the letter $H$ has been replaced by the letter $E$ for the gravitational binding energy having established their identity for geodesic observers in eqn. \ref{InterpretationQLE}. 
This relation was proposed by Penrose to understand gravity induced quantum state reduction \cite{Penrose:1996cv} and has been used in models of consciousness 
\cite{doi:10.1098/rsta.1998.0254}. Since the gravitational field energy is evaluated on the boundary this relation is inherently non-local.  
The special meaning of time emerges from the Hamiltonian acting as the operator of unit time translation. What time is depends on the choice of the boundary $\partial M$ which
is decomposed into $^3 B$ and $\Sigma$ at $t_i$ and $t_f$, respectively. The unit vectors of the latter determine what time is.
Because of the lapse being equal to $1!$ in the chosen gauge there is no distinction between proper time and coordinate time.
In the next example we derive an ordinary uncertainty relation unrelated to energy and time. 

\section{$(K)^2-K_{\mu \nu} K^{\mu \nu}=0$}
In the same way we decomposed the Hamiltonian into a stress-energy and a quasi-local energy term we proceed with the action. 
Starting from eqn. \ref{action2} and inserting eqn. \ref{RKKK2} we obtain with the condition $(K)^2-K_{\mu \nu} K^{\mu \nu}=0$
\begin{eqnarray}
\nonumber
S = \frac{1}{\kappa} \int_M d^4 x N \sqrt{h}  u^{\mu} u^{\nu} G_{\mu \nu}  \\
- \frac{1}{\kappa} \int_{^3 B} d^3 x N \sqrt{\sigma} k_1 - S_0[\gamma_{\mu \nu}] + S_m
\label{action3}
\end{eqnarray}

\section{Computation of the QLE for the modified Vaidya metric}

Continuing with eqn. \ref{action3} we derive the action of a modified Vaidya metric which has been used in the study of black hole radiance and which is given by

\begin{eqnarray}
\nonumber
ds^2 = & - & F(\nu,r,\theta) d \nu^2 + 2 d \nu dr 
\\
& + & \psi^2(r) \left ( d \theta^2 + \sin^2 \theta d \phi^2 \right )
\label{metric}
\end{eqnarray}
The fields $F$ and $\psi$ will be related by $F = 1 - 2m/\psi(r)$ later on so the problem will depend on one function to be varied only. Depending on circumstances this may or may not
results in sufficient degrees of freedom. With the unit vectors

\begin{eqnarray}
u^\mu = \frac{1}{\sqrt{F(\nu,r,\theta)}} \delta_\nu^\mu 
\end{eqnarray}
and
\begin{eqnarray}
n^\mu = u^{\mu} + \sqrt{F(\nu,r,\theta)} \delta_r^\mu
\end{eqnarray}
which satisfy the conditions $u_{\mu}u^{\mu}=-1$, $n_{\mu}n^{\mu}=1$ and $u_{\mu}n^{\mu}=0$ we obtain for the QLE related surface integral
\begin{eqnarray}
\nonumber
- \frac{1}{\kappa} \int_{^3 B} d^3 x N \sqrt{\sigma} k_1 =  \\ \nonumber
\frac{2}{\kappa} \int_{^3 B} d^3 x F(\nu,r,\theta) \sin \theta \psi(r) \cdot \frac{d}{dr} \psi(r)
\\
\label{vaidyaboundary}
\end{eqnarray}

The gauge condition is met and evaluates to zero if $F(\nu,r,\theta)=F(r)$. 
\begin{eqnarray}
K_{\mu \nu} K^{\mu \nu} - (K)^2 = \frac{\left [ \frac{\partial}{\partial \theta} F(\nu,r,\theta) \right ]^2}{2 F^2(\nu,r,\theta) \psi^2(r)}
\label{Kvaidya}
\end{eqnarray}
Furthermore, for the bulk term
\begin{eqnarray}
\nonumber
\frac{1}{\kappa} \sqrt{-g} u^{\mu} u^{\nu} G_{\mu \nu} = \\ \nonumber
\frac{\sin \theta}{4 \pi} \left ( -F \psi \psi^{\prime \prime} - \frac{1}{2} F \left ( \psi^{\prime} \right )^2 - \frac{1}{2} F^{\prime} \psi  \psi^{\prime} + \frac{1}{2} \right )
 \\
\end{eqnarray}
The term containing the second derivative can be integrated by parts. The emerging boundary term is canceled by the boundary term in the action eqn. \ref{action3} now expressed in terms of the QLE given by eqn. \ref{vaidyaboundary} which is a known property of the boundary term. 
\begin{eqnarray}
S + S_0 = \frac{1}{\kappa} \int_M d^4 x \sin \theta \left ( F   \left ( \psi^{\prime} \right )^2 + \psi F^{\prime}  \psi^{\prime} + 1 \right )
\end{eqnarray}

To lowest order in the vicinity of the event horizon we are left with
\begin{eqnarray}
S+S_0 =   \frac{1}{\kappa} \int_M d^4 x \sin \theta \left [ 1 + (\psi_{,r})^2  \right ]
\end{eqnarray}
where $\psi(r) \approx r \approx 2m$
in accordance with \cite{York:2005kn}. The extrinsic curvature terms eqn. \ref{Kvaidya} vanish as desired, so the contribution to the bulk term is due to
the stress-energy tensor alone, cf. eqn. \ref{RKKK2}. Combined with the boundary term eqn. \ref{vaidyaboundary} we obtain an action which is quadratic in the
first derivative of the field function plus a term independent of the field which could be absorbed into $S_0$ but has no impact on the classical equations
of motion. There are no second or higher powers of the field indicating that the field is essentially free and massless this gauge.

Note that $l^{\mu} = n^{\mu} - u^{\mu}$ is a null vector because the $g_{rr}$-component vanishes. 
The coordinate system leads to an effective separation of the dynamical and the gravitational degrees of freedom, and
the presence of only one term quadratic in the first derivative of the field allows for a simple quantization of this metric \cite{York:2005kn,York:1983zb}. 
Intermediate results can be found in the appendix. 

Comparing the final action with the action of a free particle in quantum mechanics we interpret $\psi$ and $\psi_{,r}$ as conjugate variables suggesting
the uncertainty relation
\begin{eqnarray}
\Delta \psi \Delta \psi_{,r} \ge c \cdot \frac{\hbar}{2}
\end{eqnarray}
where the factor $c$ is due to the integrations over $\nu$, $\theta$ and $\phi$, respectively. To lowest order this relation is equivalent to $\Delta g_{\theta \theta} \Delta g_{\theta \theta,r} \ge \tilde c \cdot \hbar / 2 $ 
which has been used in the understanding of black hole radiance
\cite{York:2005kn,York:1983zb}. 

\begin{widetext}

\begin{table}
\begin{tabular}{|r|c|c|c|c|c|} \hline
Manifold	& (Induced)	&	Covariant 	&	Unit normal 	&	Intrinsic	&	Extrinsic  \\ 
	&	metric	&	derivative	&	vector		&	curvature	&	curvature \\ \hline
$M$		&	$g_{\mu \nu}$ &	$\nabla_{\mu}$ &	&	$\mathcal{R}_{\mu \nu \rho \sigma}$ &	\\
$\Sigma$		&	$h_{ij}$	&	$D_i$	& $u_{\mu}$	&	$R_{ijkl}$ & $K_{ij}$ \\
	&	$h_{\mu \nu}=g_{\mu \nu}+u_\mu u_\nu$ & $D_\mu t^\nu = h^\alpha_\mu h^\nu_\beta \nabla_\alpha t^\beta$ & & & $K_{\mu \nu} = - h_{\mu}^{\lambda} \nabla_{\lambda} u_{\nu}$ \\ 
$^3 B$ 		&	$\gamma_{\mu \nu}$ & $\mathcal{D}_{i}$ &	$n_{\mu}$		&	&	$\Theta_{\mu \nu}$ \\
			&	$\gamma_{\mu \nu}=g_{\mu \nu}-n_\mu n_\nu$ &   $\mathcal{D}_\mu t^\nu = \gamma^\alpha_\mu \gamma^\nu_\beta \nabla_\alpha t^\beta$ & &  & $\Theta_{\mu \nu}=-\gamma_{\mu}^{\lambda} \nabla_{\lambda} n_{\nu}$ \\
$B$		&	$\sigma_{\mu \nu}$	&	&	$n_{\mu}$ &	&	$k_{\mu \nu}$ \\
		&	$\sigma_{\mu \nu} =g_{\mu \nu} - n_\mu n_\nu + u_\mu u_\nu$ & & & & $k_{\mu \nu}=-\sigma_{\mu}^{\lambda} D_{\lambda} n_{\nu}$ \\
\hline
\end{tabular}
\caption{Summary of notation. Note the change of notation for the intrinsic curvatures deviating from past works. }
\label{notation}
\end{table}

\end{widetext} 

\section{Alternative Constraints}
For a simple interpretation of the terms in the Hamiltonian and in the action we prefer terms which represent energies. Possible gauge choices which may accomplish this include
$R+K_{\mu \nu} K^{\mu \nu} - (K)^2=0$ which would leave the action with the quasi-local energy term only or $K_{\mu \nu} K^{\mu \nu} - (K)^2 = A u^{\mu} u^{\nu} G_{\mu \nu}$ with $A$ being an unknown constant. Enforcing these conditions may be very hard in practice if the constraint equation can be solved at all. If possible the action would still consist of a bulk and a surface term with both of them representing energies. In the former case the action contains a surface term only. 

\section{Conclusions}
Standard manipulations in differential geometry allowed us to rewrite the bulk term in the action and in the Hamiltonian as pure stress-energy. This was achieved by employing
the geodesic gauge and the gauge condition $K_{\mu \nu} K^{\mu \nu} - (K)^2=0$, respectively. This simplifies the interpretation of the different terms because the bulk term
and the boundary term can be interpreted as either stress-energy or gravitational field energy.  Other gauge conditions like 
$K_{\mu \nu} K^{\mu \nu} - (K)^2 = A u^{\mu} u^{\nu} G_{\mu \nu}$ with $A$ being an unknown constant are possible, but presently we find no use in such generality. 
The used gauges remove coordinate effects leaving particularly simple forms for the action consisting of a square of a first derivative term only. 
Corresponding uncertainty relations have been derived. 

\acknowledgments
The author would like to acknowledge insightful discussion with James W. York, Jr. 
Partial support for this project provided by the National Science Foundation under contract \# PHY-0714648 is also acknowledged. 

\appendix
\section{Intermediate Results}
\begin{eqnarray*}
\tau^{\nu \nu} = - \frac{2 \psi_{,r}}{\kappa \psi \sqrt{F}}  \\
\tau^{\theta \theta} = \frac{2F^2 \psi_{,r} + FF_{,r} \psi - \psi F_{, \nu}}{2 \kappa \psi^3 F^{3/2}} \\
\tau^{\nu \theta} = \tau^{\theta \nu} = \frac{F_{, \theta}}{2 \kappa \psi^2 F^{3/2}} \\
\tau^{\phi \phi} = \frac{2F^2 \psi_{,r} + \psi F F_{,r}-\psi F_{, \nu}}{2 \kappa \psi^3 F^{3/2} \sin^2 \theta} \\
K_{rr} = \frac{1}{2} \frac{F_{, \nu}}{F^{5/2}} \\
K_{\theta r} = \frac{1}{2} \frac{F_{, \theta}}{F^{3/2}} = - K_{r \theta}\\
G_{\nu \nu} = - \frac{1}{2} \psi^{-2} \left [ 4 \psi F^2 \psi_{,rr} + 2F_{,r} \psi_{,r} F \psi  \right . \\
\left . + 2 F^2 \left ( \psi_{,r} \right )^2 + 2 F_{, \nu} \psi_{,r} \psi - 2F - F_{, \theta \theta} - F_{, \theta} \cot \theta \right ] \\
G_{rr} = - \frac{2 \psi_{,rr}}{\psi} \\
G_{\theta \theta} = \frac{\psi}{2} \left [ \psi F_{,rr} + 2 F \psi_{,rr} + 2 F_{,r} \psi_{,r} \right ] \\
G_{\phi \phi} = \frac{\psi}{2} \sin^2 \theta \left [ \psi F_{,rr} + 2 F \psi_{,rr} + 2 F_{,r} \psi_{,r} \right ] \\
G_{r \nu} = \psi^{-2} \left [ 2F \psi \psi_{,rr} + F_{,r} \psi \psi_{,r} + F \left ( \psi_{,r} \right )^2 - 1 \right ] \\
G_{\theta \nu} = -\frac{1}{2} F_{,r \theta} \\
2 \kappa u^{\mu} u^{\nu} T_{\mu \nu} = - \frac{1}{F \psi^2 } \left [ 4 \psi F^2 \psi_{,rr} + 2 F_{,r} \psi_{,r} F \psi + 2 F^2 \left ( \psi_{,r} \right )^2 + \right . \\
\left . 2 F_{,\nu} \psi_{,r} \psi - 2 F - F_{, \theta \theta} - \cot \theta F_{, \theta} \right ]
\end{eqnarray*}

\bibliographystyle{apsrev4-1}
\bibliography{bib}

\end{document}